\documentclass[sigchi]{acmart}

\usepackage{booktabs} 

\usepackage{subfig}
\usepackage{balance}

\AtBeginDocument{%
  \providecommand\BibTeX{{%
    \normalfont B\kern-0.5em{\scshape i\kern-0.25em b}\kern-0.8em\TeX}}}

\setcopyright{rightsretained}
\copyrightyear{2019} 
\acmYear{2019} 
\acmConference[ICMI '19]{2019 International Conference on Multimodal Interaction}{October 14-- 18, 2019}{Suzhou, China}
\acmBooktitle{2019 International Conference on Multimodal Interaction (ICMI '19), October 14--18, 2019, Suzhou, China}\acmDOI{10.1145/3340555.3353741}
\acmISBN{978-1-4503-6860-5/19/10}

\begin{document}

\title{Comparing Pedestrian Navigation Methods in Virtual Reality and Real Life}

\author{Gian-Luca Savino}
\orcid{0000-0002-1233-234X}
\affiliation{\institution{University of Bremen}}
\email{gsavino@uni-bremen.de}

\author{Niklas Emanuel}
\affiliation{\institution{University of Bremen}}
\email{emanuel@uni-bremen.de}

\author{Steven Kowalzik}
\affiliation{\institution{University of Bremen}}
\email{ste_kow@uni-bremen.de}

\author{Felix A. Kroll}
\orcid{0000-0003-2195-3903}
\affiliation{\institution{University of Bremen}}
\email{fe_kr@uni-bremen.de}

\author{Marvin C. Lange}
\affiliation{\institution{University of Bremen}}
\email{marvin4@uni-bremen.de}

\author{Matthis Laudan}
\affiliation{\institution{University of Bremen}}
\email{laudan@uni-bremen.de}

\author{Rieke Leder}
\affiliation{\institution{University of Bremen}}
\email{rleder@uni-bremen.de}

\author{Zhanhua Liang}
\affiliation{\institution{University of Bremen}}
\email{zhanhua1@uni-bremen.de}

\author{Dayana Markhabayeva}
\affiliation{\institution{University of Bremen}}
\email{dayana1@uni-bremen.de}

\author{Martin Schmei{\ss}er}
\affiliation{\institution{University of Bremen}}
\email{s_h6krpo@uni-bremen.de}

\author{Nicolai Sch\"utz}
\affiliation{\institution{University of Bremen}}
\email{s_ighm5k@uni-bremen.de}

\author{Carolin Stellmacher}
\affiliation{\institution{University of Bremen}}
\email{cstellma@uni-bremen.de}

\author{Zihe Xu}
\affiliation{\institution{University of Bremen}}
\email{zihe@uni-bremen.de}

\author{Kerstin Bub}
\affiliation{\institution{University of Bremen}}
\email{kerstin.bub@uni-bremen.de}

\author{Thorsten Kluss}
\affiliation{\institution{University of Bremen}}
\email{tox@uni-bremen.de}

\author{Jaime Maldonado}
\affiliation{\institution{University of Bremen}}
\email{jmaldonado@uni-bremen.de}

\author{Ernst Kruijff}
\orcid{0000-0003-1625-0955}
\affiliation{\institution{University of Applied Sciences Bonn-Rhein-Sieg}}
\email{ernst.kruijff@h-brs.de}

\author{Johannes Sch\"oning}
\orcid{0000-0002-8823-4607}
\affiliation{\institution{University of Bremen}}
\email{schoening@uni-bremen.de}

\renewcommand{\shortauthors}{Savino et al.}

\begin{abstract}
  Mobile navigation apps are among the most used mobile applications and are often used as a baseline to evaluate new mobile navigation technologies in field studies. As field studies often introduce external factors that are hard to control for, we investigate how pedestrian navigation methods can be evaluated in virtual reality (VR). We present a study comparing navigation methods in real life (RL) and VR to evaluate if VR environments are a viable alternative to RL environments when it comes to testing these. In a series of studies, participants navigated a real and a virtual environment using a paper map and a navigation app on a smartphone. We measured the differences in navigation performance, task load and spatial knowledge acquisition between RL and VR. From these we formulate guidelines for the improvement of pedestrian navigation systems in VR like improved legibility for small screen devices. We furthermore discuss appropriate low-cost and low-space VR-locomotion techniques and discuss more controllable locomotion techniques.
\end{abstract}




\maketitle

\begin{figure}[t!]
\captionsetup[subfloat]{labelformat=empty}
\centering
\subfloat[(a)]{
\includegraphics[width=0.40\linewidth]{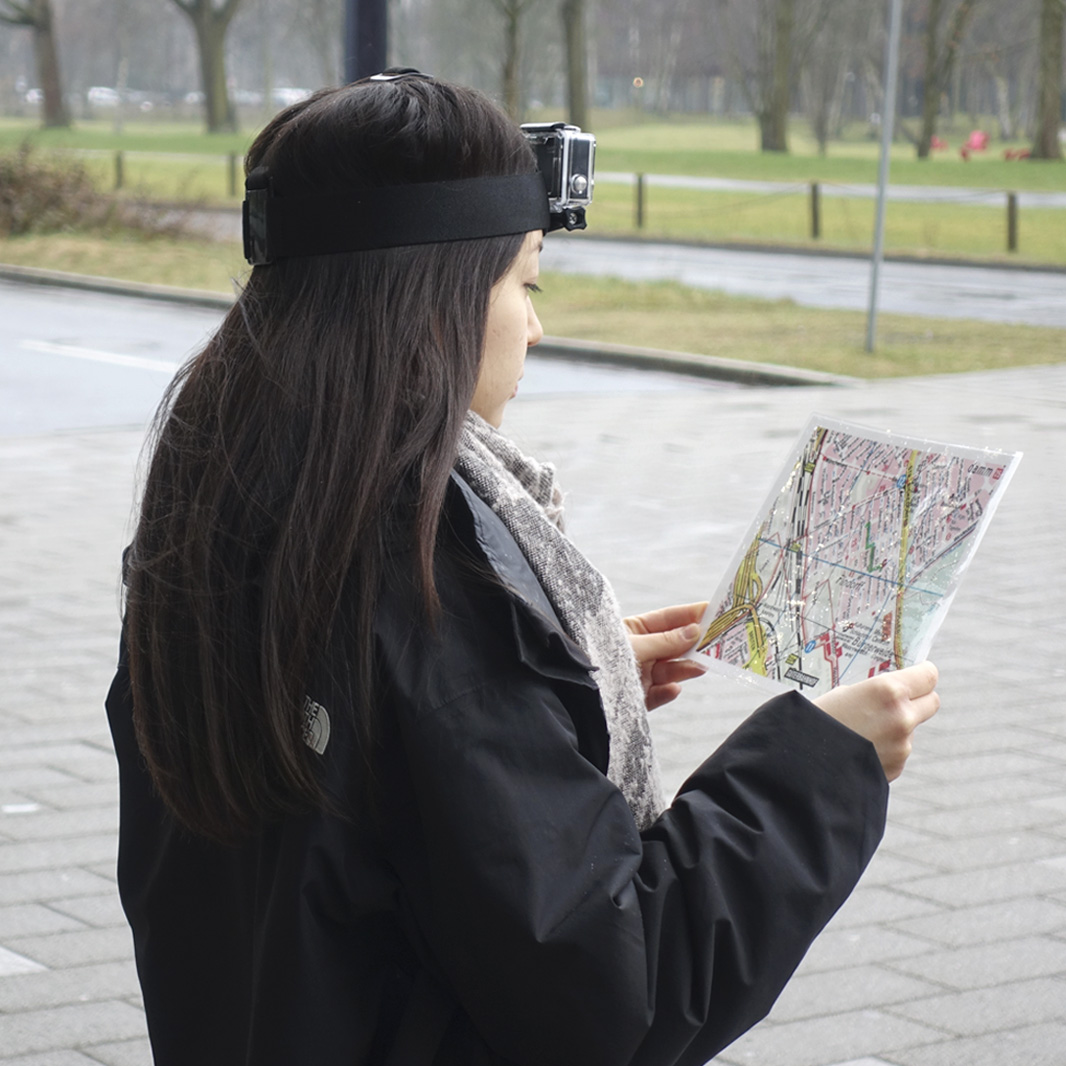}
}
\quad
\subfloat[(b)]{
\includegraphics[width=0.40\linewidth]{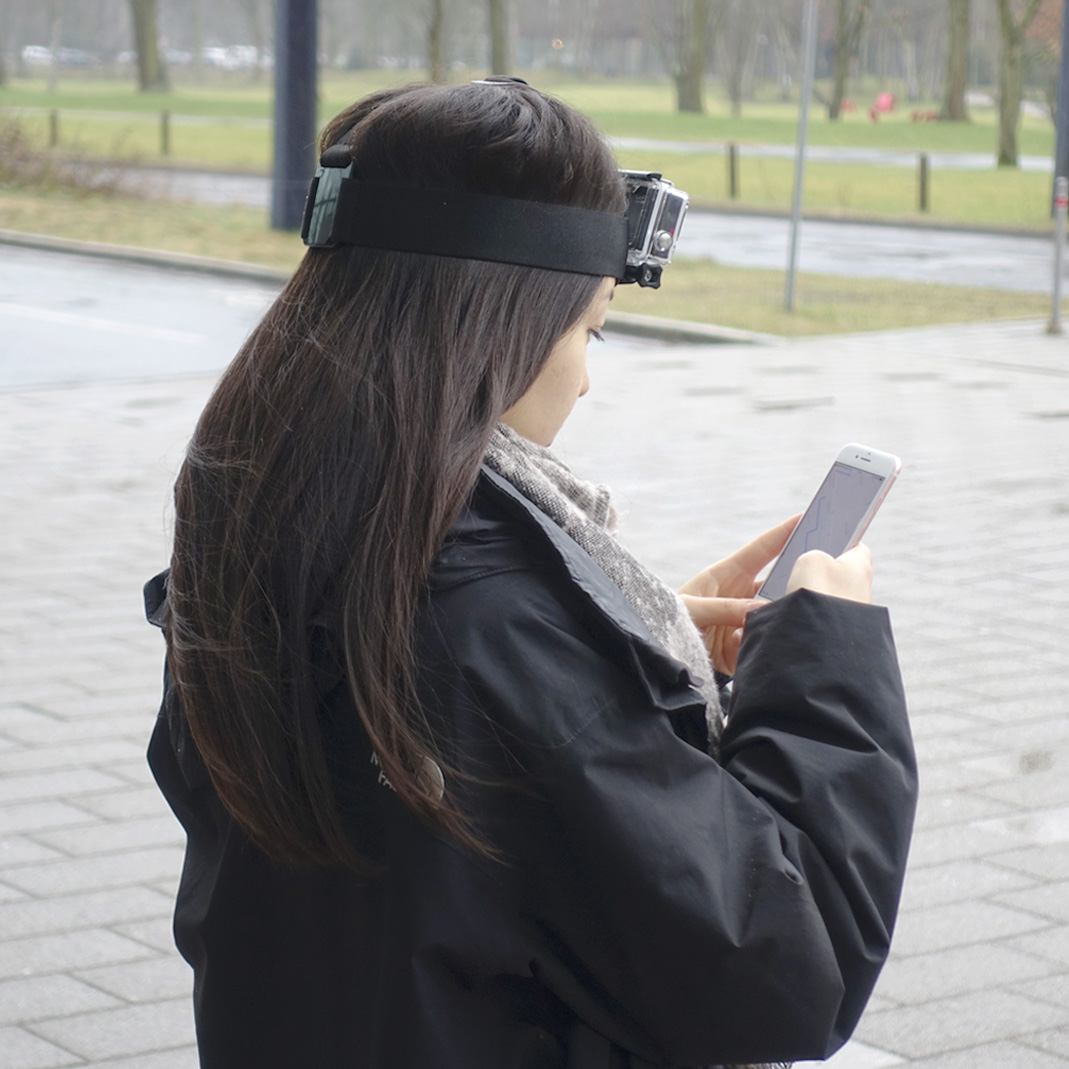}
}
\\
\subfloat[(c)]{
\includegraphics[width=0.40\linewidth]{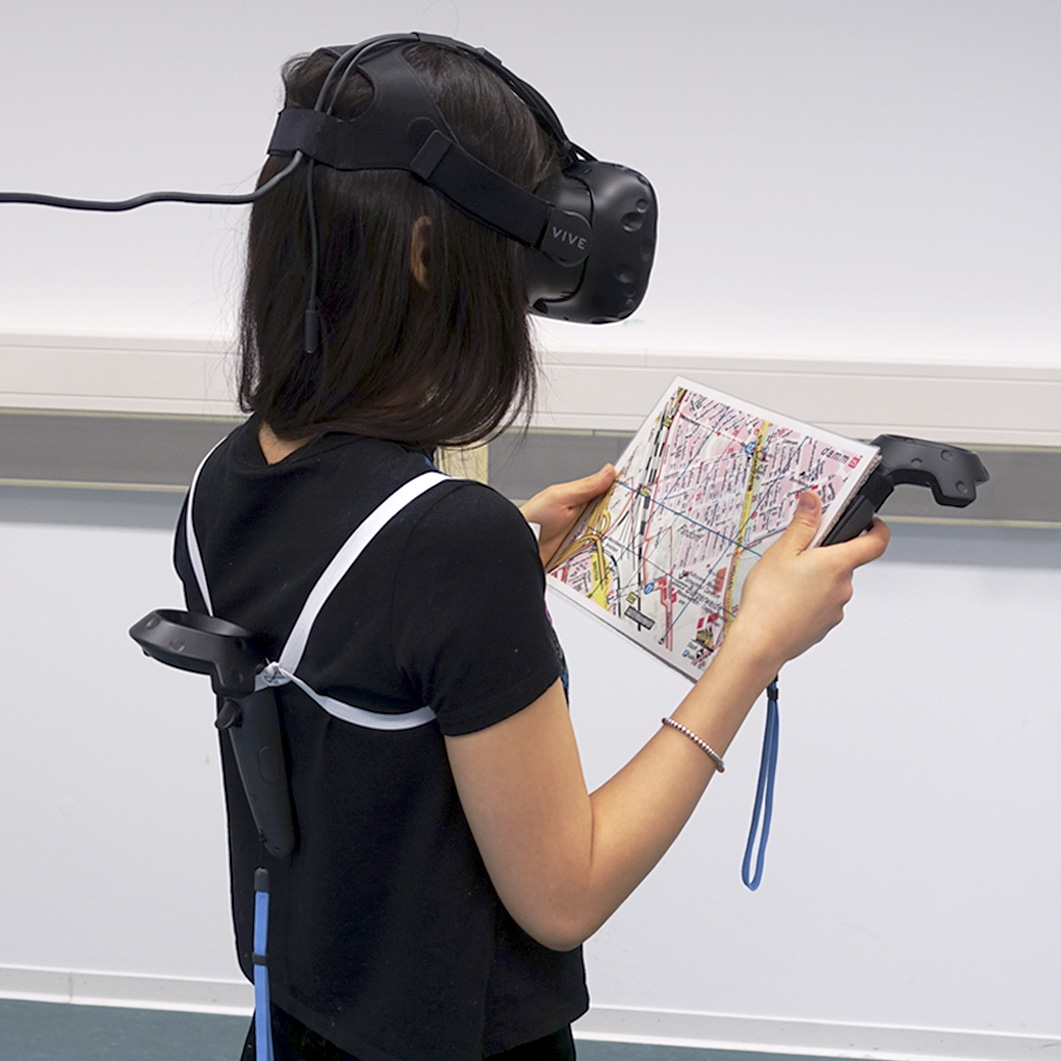}
}
\quad
\subfloat[(d)]{
\includegraphics[width=0.40\linewidth]{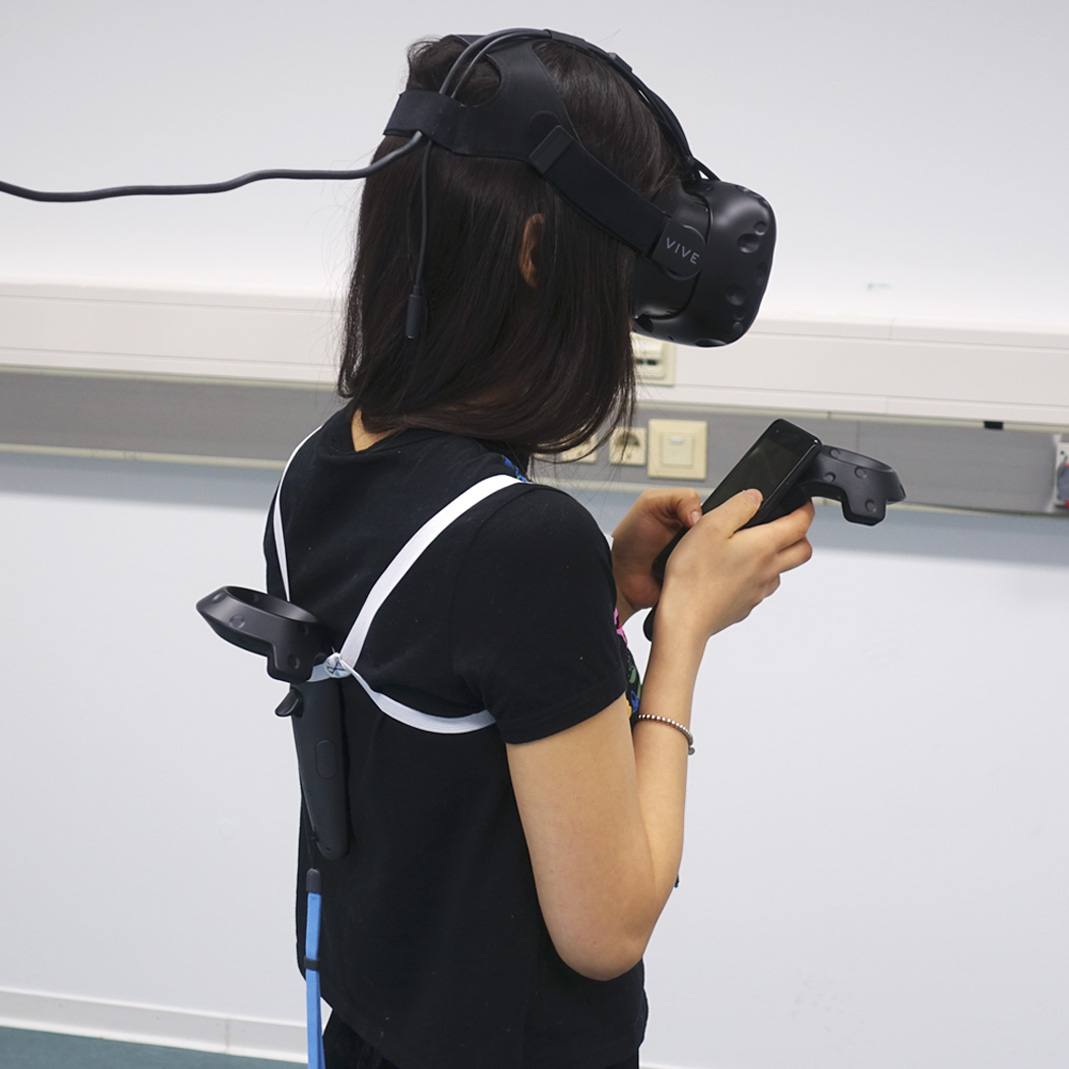}
}
\caption{User study conditions: Paper map in RL (a), smartphone in RL (b), paper map attached to a VIVE controller for VR (c), and smartphone attached to a VIVE controller for VR (d).}~\label{fig:navigation_methods}
\end{figure}

\section{Introduction \& Motivation}
Mobile navigation applications like \emph{Google Maps} are among the most used smartphone apps~\cite{Frommer:2017}. Using mobile devices to navigate has become a very frequent task, making mobile map apps the de facto standard for navigation. Still a diverse portfolio of novel navigation methods and multimodal interfaces is continuously developed, such as tactile belts~\cite{Pielot:2010}, wearable displays~\cite{Dancu:2015}, drones for displaying spatial information~\cite{Schneegass:2014} or smartwatch-navigation interfaces~\cite{Wenig:2015}. Typically, these methods have been compared against a baseline technique (e.g. paper map or mobile map app) in a field study. However, those field studies in real life environments are hard to compare, as they are typically conducted in different cities with different city structures, different landmarks and different weather and traffic conditions. 
Therefore, the goal of this paper is to evaluate if VR environments are a viable alternative to RL environments when evaluating  navigation interfaces and methods. We highlight problems and challenges of this approach and formulate guidelines on how to evaluate navigation methods in VR and RL. Our work builds upon a large body of research on evaluating  navigation methods and devices in the lab~\cite{Kjeldskov:2004, Farrel:2003, Cliburn:2007} and especially in VR~\cite{Richardson:1999, Wallet:2008, Feldstein:2016}. We extend this literature by comparing the effectiveness of this approach and discuss the influence VR has on the two most common navigation methods (paper map and smartphone). If VR environments were to mimic most of the qualities of a RL study, we could potentially eliminate external factors common in RL field studies. A first step towards this is testing the two most common navigation methods under similar conditions in VR and RL to answer the following research questions: (RQ1) How does the locomotion, the VR environment and the navigation method influence participants navigation performance? And (RQ2) how can we improve upon these factors and bring VR navigation and RL navigation closer together? We created a VR environment that is geographically identical to Findorff a part of Bremen in Germany and compared the navigation performance using a paper map and a smartphone with a mobile map app. Participants performed the same navigation tasks in RL and VR (see figure \ref{fig:navigation_methods}). To test participants in both environments the navigation performance (defined by time and number of errors), task load and the spatial knowledge acquisition were measured. To assure a natural and immersive VR experience we evaluated four different VR locomotion techniques and compared them to walking in RL prior to the main study. 

\subsection{Contributions}
Overall, our results show that participants navigating in VR show a different performance from participants in the RL environment. However, as related literature suggests we see promising results for the spatial knowledge acquisition. Therefore we provide insights and formulate guidelines for (but not limited to) the following findings: In VR participants performed better with the smartphone than with the paper map. In RL participants performed better with the paper map than with the smartphone. Individual differences increase in VR. Participants experience higher task load in VR than in RL regardless of the navigation method. Participants acquire the same spatial knowledge in VR and RL in a route recall test. In VR participants quickly adapt to the virtual setting and try to exploit the VR locomotion technique.

\section{Related work}
There is a large body of related work studying various aspects of pedestrian navigation. Researchers are interested in navigation strategies and performance~\cite{Pielot:2010}, task load or spatial knowledge acquisition during navigation~\cite{Burigat:2011, Szymczak:2012, Aslan:2006, Wallet:2008}. There is also a long tradition of building novel and multimodal navigation prototypes~\cite{Dancu:2015,Schirmer:2015} and comparing them to existing techniques such as turn-by-turn navigation on a mobile device or paper map, using field studies. But evaluating navigation tasks in laboratory settings is also becoming more common. Virtual environments (VEs) with different levels of immersion and locomotion techniques have been a widely used tool in those studies. Cliburn~\cite{Cliburn:2007} for example used three large projection screens oriented at 120 degree angles to conduct an experiment exploring the use of dynamically placed landmarks as navigation aids. Subjects used a gamepad to move and rotate through the VE and interact with the system. For pedestrian street crossing simulations, Deb et al.~\cite{Deb:2017} and Feldstein et al.~\cite{Feldstein:2016} conducted studies in VR using low cost HMDs with real walking in a small tracking space. Large screens, projections or HMDs are ways to improve the visual immersion \cite{Feldstein:2016,Deb:2017,Lehsing:2016,Simpson:2003,Maffei:2014} that before was restricted to regular computer screens. Nevertheless there was already extensive research being conducted on navigation that used the benefit of a lab environment even with basic computer monitors. 

\subsection{Spatial Learning in Virtual Environments}
Cushman et al.~\cite{Cushman:2008} relied on the flexibility and safety of the laboratory in order to investigate whether navigational deficits caused by cognitive ageing and Alzheimer's disease observed in RL could be observed in VR navigation. For this purpose, navigation in RL and VR was compared. In the RL experiment, participants sat in a wheelchair and were carried by an experimenter through an indoor environment following a fixed route. In VR participants were sitting in front of a display and navigated through the simulated environment. Their results show that there is a close correlation between RL and VR navigational deficits and they conclude that virtual navigation is well suited for conducting experiments in this domain. Research also shows that virtual environments can successfully be used to learn spatial information. Wallet et al.~\cite{Wallet:2008} have shown that spatial knowledge can be transferred successfully from a VE to the real world when using action based learning. Participants navigated two routes with three different display modes: (1) passive VE (with a route recorded), (2) active VE with joystick and (3) the real environment (the participant actually travelled the route by following instructions). The joystick outperformed the passive condition in both way-finding and sketch mapping tasks. This finding is supported by the results of Richardson et al. ~\cite{Richardson:1999}. They compared the spatial knowledge acquisition of maps, real- and virtual-indoor environments by letting the participants learn in one of these three conditions and testing their route distance and direction estimates to specific landmarks in the real environment afterwards. Their results point out that all three learning conditions performed similarly on remembering the layout of landmarks on a single floor, though the VE learners achieved the lowest performance. A similar result was found by Bliss et al. by showing that firefighters learning to navigate through a building in VR were almost as successful as the one learning in RL~\cite{Bliss1997}. The above literature shows that lab studies on navigation have been successfully conducted using everything from computer monitors to HMDs and that knowledge and experiences transfer partially from virtual to real environments. Based on this premise we designed our pedestrian simulator in VR and test it against RL conditions. 

\section{Pre-Study: VR Locomotion Techniques}
An existing issue of studies using virtual environments is the break of realism and immersion due to  VR locomotion techniques~\cite{Delikostidis:2015}. Therefore, the goal of the preliminary study was to evaluate different locomotion techniques in terms of pace, naturalness and comfort. Unrealistic VR motion disturbs the experience in terms of naturalness and immersion, while users can also suffer from severe motion sickness \cite{Bles:1998} when exposed to visual-proprioceptive conflicts.  
A reasonable solution is to match the tracking space to the size of the VE, but is often not feasible because VEs are larger in size, as compared to the space available in laboratories. When the size of the VE exceeds the available space in the laboratory, people have used various complex (hardware) setups to enable locomotion in VR. Examples are the \emph{VirtuSphere} \cite{Kluss:2015}, \emph{Suspended Walking} \cite{Walther-Franks:2013}, \emph{Omnidirectional Treadmills} \cite{Darken:1997} or \emph{Redirected Walking} \cite{Suma:2015}. While those setups offer somehow natural locomotion, they make results hard to compare across different user studies. Therefore we focused on comparing locomotion techniques which do not require additional hardware and have low space and low cost requirements to ensure a wide range of applications and thus comparability. In our pre-study we compare walking in a real environment with \emph{100\,ms Warp}, \emph{Avatar Warp}, \emph{Walking-in-Place} and \emph{Freeze Rotation} in a virtual environment,
which come at a low price and need a minimal amount of space (3m x 3m).
WARP is a sub-type of the arch teleportation location technique.  Arch teleportation is the standard locomotion technique in the \emph{Steam VR} home app. 
AVATAR is also a sub-type of arch teleportation. It requires the participants to choose a target destination, which is then approached by a human avatar. 
When confirming, the participant is teleported to the avatars position. Walking-in-Place (WIP) measures the head elevation during a stepping motion.
The movement is then translated into VR motion, which enables the participant to navigate through the VE while actually walking in place. Freeze Rotation (FREEZE) is a locomotion technique in which participants walk through the VR tracking space. When reaching the border of the tracking space participants can confirm so using a button press which freezes the VR scene. They can then turn around in the tracking space and release the button. This way they have ``frozen'' the scene and therefore not changed the orientation in VR but can now walk through the VR tracking space again.

\subsection{Participants \& Procedure}
We recruited 16 participants (11 male and 5 female) with an average age of 25 (min: 20, max: 30). 
Every participant navigated a 440m path in RL and VR with the different locomotion techniques. The study used a within-subject design and the conditions were counterbalanced. 
The participants were provided with instructional paper cards that showed the navigation instructions (``left'', ``right'' or ``straight'') at the 24 decision points. The participants were asked to follow the instruction each time they encountered a decision point  simulating a turn-by-turn style navigation approach. In RL an experimenter followed them to measure the duration, count the navigation errors and number of stops they made.

\subsection{Results \& Analysis}

\begin{figure}[t!]
\centering
  \includegraphics[width=\columnwidth]{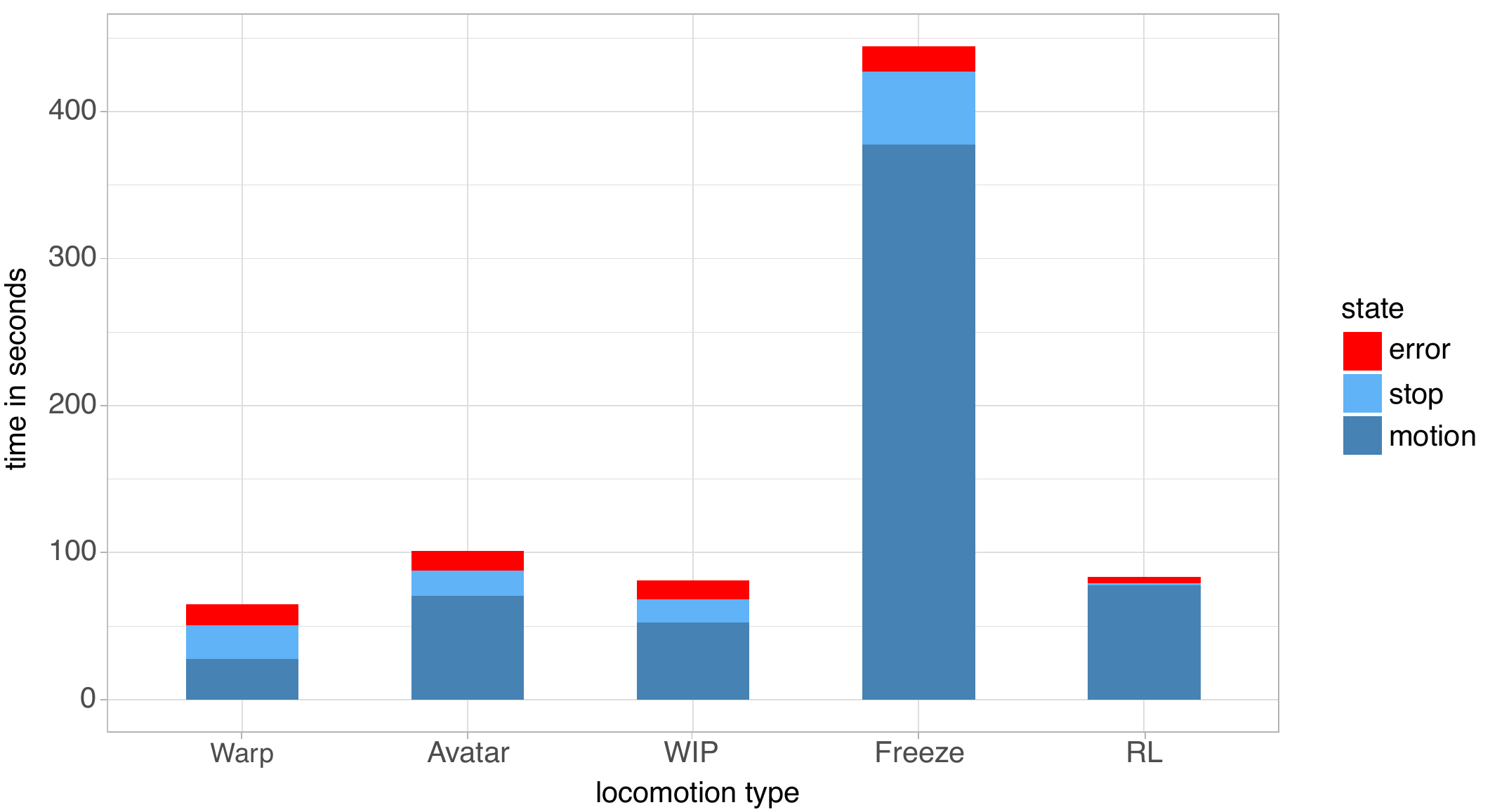}
  \caption{The average time spent in motion, stopping + standing and error correction per segment (110m).}
\label{fig:completion_time}
\end{figure}

We conducted a one-way repeated measure analysis of variance (ANOVA) for completion time measurements. It revealed a significant effect of the locomotion technique on completion time performance ( F(4,75) = 186.4, p < .001). For post-hoc testing we used Sidak-Correction. Pairwise t-testing with repeated measures of the completion time revealed significant differences between FREEZE (M = 444.65, SD = 95.30) and all other locomotion techniques respectively. Additionally, WARP (M = 65.06, SD = 29.73) and AVATAR (M = 101.03, SD = 20.49) showed significant differences (t(15) = -5.83, p < .001). When comparing the means WIP (M = 80.98, SD = 27.98) was the closest to RL (M = 83.27, SD = 12.51), while WARP, AVATAR as well as FREEZE showed larger differences. Each participant made at least 1 navigation error and more than 8 stops  with an average of 2.44 errors (SD = 1.09) and 16 stops (SD = 7.42). The main reasons for stopping and standing were orientation and instruction comprehension. Pairwise Wilcoxon-Signed-Rank tests showed that participants made significantly more stops when moving in each VR locomotion technique compared to RL (M = 0.45, SD = 0.44). In terms of the times spent for resolving errors we found no significant effect between the locomotion techniques. In contrast to the time participants stopped, a non-parametric Friedman-Test showed a high significant impact (Chi-Square = 39.25, p < .001) of the locomotion technique on the stop time. In all VR locomotion techniques participants spent significantly more time standing when compared to RL. From all VR locomotion techniques WIP has the lowest stop rate, standing time and error resolving time. Answers regarding the perceived naturalness and comfort for each locomotion technique were collected using a 6-level forced-choice Likert-scale (0 = not close to RL walking, 5 = very close to RL walking). Post-hoc pairwise Wilcoxon-Signed-Rank tests showed that WIP (M = 3.19, SD = 0.98) was significantly higher rated in terms of naturalness when compared to all other locomotion techniques respectively. WARP (M = 0.94, SD = 1.29) was rated the least natural, followed by AVATAR (M = 1.50, SD = 1.21) and FREEZE (M = 1.69, SD = 1.62). A one-way repeated measures ANOVA on the responses about comfort also showed a significant effect between the locomotion techniques (F(3,60) = 6.01, p = .001). Post-hoc pairwise t-tests with repeated measures revealed that FREEZE (M = 1.50, SD = 1.55) has been considered significantly less comfortable by participants when compared to all other locomotion techniques. The most comfortable locomotion technique was WARP (M = 3.31, SD = 1.49), which was slightly better than WIP (M = 3.00, SD = 1.10) and AVATAR (M = 2.94, SD = 1.06).

We asked participants to rank the perceived accuracy of the pace for each locomotion technique compared to RL walking on a 5 level Likert-scale (much slower, slower, accurate, faster, much faster). Medians for each locomotion technique are as follows: WARP (much faster), Avatar (accurate), WIP (accurate), FREEZE (much slower). When asked to rank the locomotion techniques according to their own assessment half of the participants ranked WARP the best, while 5 participants preferred WIP.

\subsection{Discussion \& Implications}
WIP had the closest mean completion time compared to RL walking (see figure \ref{fig:completion_time}). 
In terms of naturalness WIP performed significantly better than the other locomotion techniques, which might be an effect of a more comprehensive full body motion involvement. The only other locomotion technique including a similar body movement was FREEZE, which performed poorly in pace and comfort.
Although WARP had the highest level of comfort, it was significantly faster than RL, participants considered it very unnatural and they spent a relatively long time standing and resolving errors (58\% of the total time). This highlights the effect of unrealistic and comfortable movement on effective and natural navigation. In both WARP and AVATAR, participants sometimes simply missed a decision point, because they teleported too far, resulting in navigation errors. WIP and FREEZE offered a more continuous motion, but with FREEZE being unacceptably uncomfortable. Therefore we decided to use WIP in the main study as it provides a good  balance between pace, naturalness and comfort.

\section{Main Study: VR Environment \& Navigation Methods}
The main study investigated participants navigation behaviour with two different methods: A paper map and a mobile map application on a smartphone. The aim was to test these under similar conditions in VR and RL to answer our research questions. (RQ1)  How does the locomotion, the VR environment and the navigation method influence participants navigation performance? And (RQ2) how can we improve upon these factors and bring VR navigation and RL navigation closer together?

The study was conducted using a within-subject design meaning each participant tested every navigation method. This results in the following experiment conditions: Navigating in RL with a paper map (RL PM), navigating  in RL with a smartphone (RL SP),  navigating  in VR with a paper map (VR PM) and navigating  in VR with a smartphone (VR SP). In each condition the participants performed the navigation task of following a predefined route. Instructions for this were provided by displaying the route on both the paper map and the smartphone respectively in RL and VR. The smartphone additionally provided participants with their current location. To avoid any spatial learning effects between the navigation tasks (conditions) we chose four different routes, which are located in a residential district and randomised the conditions. The area consists of many intersecting streets and allowed us to determine four distinct routes (approx. 420m per route), each of which had five turns.

\subsection{Measures}
This section presents the measures that were used to evaluate the navigation performance, task load, and spatial knowledge in the experiment. 


The navigation performance indicates how well participants performed according to the time they took to complete the navigation task and how many errors they made. We call the overall time participants took to complete the route, completion time (CT). It consists of the movement time (MT, time participants are actively moving) and the stop time (ST, time participants stop and stand still). When participants performed a wrong turn at a decision point, an error was counted. The error time (from performing the error until participants were back on the predefined track) was removed from the CT for all conditions.

The task load index is calculated according to the NASA TLX Paper and Pencil Package~\cite{Hart1988}.

The acquired spatial knowledge is measured by two different tests after each condition: A landmark knowledge test and a route knowledge test.
Acquired landmark knowledge is determined by a landmark recall task based on the scene recognition test~\cite{Lapeyre:2011}. For each route nine pictures are shown to the participants of which five are actually showing parts of the route, while the others are unrelated. We ask the participants to decide for each picture if they recall seeing it during the navigation task. For each correct answer they score one point, resulting in a maximum score of nine points per test. The route knowledge test includes a sketch-drawing task that requires the participant to recall the travelled route~\cite{Wallet:2008}. We ask the participants to indicate directional changes and the amount of turns by drawing the route on a blank piece of paper.

\subsection{Implementation}\label{implementation}
In the RL environment participants received either a laminated DIN A4 sized paper map (figure \ref{fig:navigation_methods} (a)) or a smartphone (iPhone 7, figure \ref{fig:navigation_methods} (b)) showing the experiment area and one of the four predefined routes. The \emph{iOS} app used the \emph{Google Maps} API for a basic map visualisation and to provide the participants with the route information and their current location. The smartphone was also used in the RL PM condition to track navigational data such as the users' location and speed data. For the VR setup we used the \emph{HTC Vive} and the \emph{SteamVR} software.
To accurately rebuild the RL environment in VR the layout is based on \emph{OpenStreetMap} data and modelled in \emph{3DS Max}. After completing the basic setup of the environment the geometry was manually refined based on \emph{Google Street View} imagery. Landmarks, street signs and other details were modelled in \emph{Blender} and added to the scene.
We use a slightly modified version of the WIP technique compared to the one used in the preliminary study. This version uses one of the VIVE controllers worn as a backpack (as depicted in figure \ref{fig:navigation_methods} (c \& d)). The controller's up and down movement was tracked to translate the participant's WIP motion into a forward motion in VR. The grip button of the other controller was used to toggle the usage of the WIP locomotion.
To simulate the natural haptic feedback of the paper map and the smartphone they were attached to the controller with velcro (see figure \ref{fig:navigation_methods} (c \& d)), depending on the condition. To match the participants preferred hand the controller could be attached on either the map's right or left side. The smartphone ran a network client application created in \emph{Unity} which allowed us to use the full multi-touch functionality of the device. The application used \emph{UNET}, \emph{Unity's} high-level networking API to communicate with the main study setup. All input was evaluated on the smartphone, while the detected gestures (e.g. touch, pan, pinch and rotate) were sent to the server. The corresponding actions were then performed on the virtual smartphone, making for a fully immersive smartphone interaction.

\subsection{Participants \& Procedure}
We recruited 16 participants (8 male and 8 female) with an average age of 24 (min: 18, max: 30) through advertisements on local online bulletin boards. People with strong visual or hand-motor system impairments were excluded. In a demographic survey 11 participants stated to have used HMDs for a few times, 4 participants never used HMDs before and one participant has been using it on a regular basis. All participants use navigation applications, varying from every day to rarely.
Depending on the starting condition for the participants, we met them either at the RL environment or at the laboratory where the VR setup was located. In RL they were guided to the respective starting point of each route. At the starting point participants received instructions for the navigation task, a remark about the tests after each task and were equipped with a \emph{GoPro} (see figure \ref{fig:navigation_methods} (a \& b)). Participants performed the task on their own and were picked up at the route's destination. After reaching the destination, the participants performed the \emph{NASA TLX} and landmark recall test on an iPad, the route recall test was done with pen and paper.
The VR part of the experiment started with a three minute introduction to the \emph{HTC VIVE} and a training for the WIP locomotion technique. Afterwards the participants were introduced to the first navigation method and had some additional time to get used to it. Afterwards they were placed into the experiment environment and performed the task. Like in the RL condition the participants had to take a route and landmark recall test as well as the NASA TLX after each run. After the experiment they were asked to fill in a demographic questionnaire also including questions about their VR experience (e.g. if they experienced simulator sickness) during the experiment. This was followed by a semi-structured interview. The interview was done in the participants' preferred language, either English or German and included questions about their perceived difference between RL and VR as well as between the navigation methods. Between the VR and RL conditions participants were escorted by car (10 minutes) to their next destination.

\section{Results \& Analysis}
Half of the participants reported having simulator sickness in VR and rated the influence on their performance with 2.5 on average (0 = no influence, 4 = strong influence). When asked about how familiar participants were with the RL environment the average answer was 2.25 (0 = not familiar at all, 5 = very familiar).

 \begin{figure}[t!]
 \centering
   \includegraphics[width=1\columnwidth]{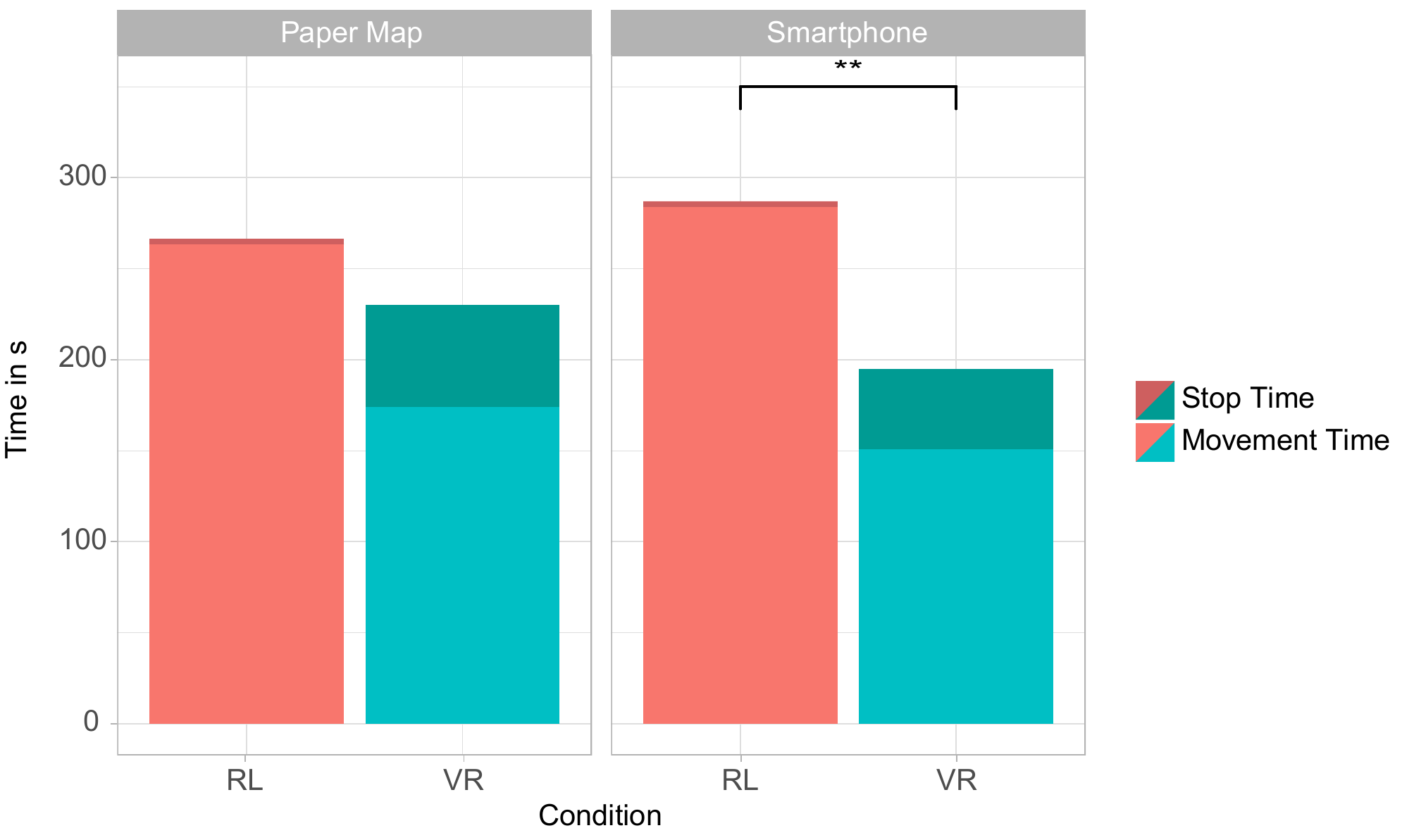}
   \caption{Average completion time for each method in each condition.}
   \label{fig:ct}
 \end{figure}
 
\subsection{Navigation Performance}
In RL the average completion time was $266.38$ seconds ($SD = 32.30$) when using a paper map and $287.12$ seconds ($SD = 44.87$) when using a smartphone. In VR the participants were faster in both the PM ($M = 230.23,\ SD = 100.67$) and SP condition ($M = 195.03,\ SD = 102.40$). The standard deviation was 2-3 times larger in VR than in RL, which indicates higher individual differences in VR. A Wilcoxon-Signed-Rank test revealed significant differences when comparing the SP conditions ($p  < .05,\ Z = -2.543$), whereas no significance was found in the PM condition ($p = 0.193,\ Z = -1.302$). Figure \ref{fig:ct} shows the mean CTs for all conditions.
\subsubsection{Movement Time}
When using the paper map participants were on average slower in RL ($M = 263.16,\ SD = 35.86$) than in VR ($M = 173.97,\ SD = 91.15$). Similar results were found in the smartphone conditions, where in RL ($M = 283.76,\ SD = 43.88$) the movement time was higher than in VR ($M =150.78,\ SD = 79.04$). Pairwise t-tests with repeated measures of the movement time revealed significant differences for both conditions (PM: $t(15) = 4.26,\ p < .001$, SP: $t(15) = 5.81,\ p < .0001$). However, in RL the movement time when using a smartphone was increased compared to the paper map condition, while in VR we found the opposite. 
\subsubsection{Stop Time}
When using the paper map in RL 4 participants stopped at least once, while the others walked continuously. The smartphone induced slightly more stops, but still five participants never stopped. In VR all participants stopped at least once, regardless of the method. Correlating to these numbers was the difference in stop time between RL and VR. A Wilcoxon-Signed-Rank test found significant differences ($p < .001,\ Z = -3.491$) when comparing the stop time in RL PM ($M = 3.22,\ SD = 6.99$) to the VR counterpart ($M = 56.26,\ SD = 37.65$). Performing the same test on the smartphone conditions also revealed significant differences ($p < .0001,\ Z = -4.170$) between RL ($M = 3.36,\ SD = 3.54$) and VR ($M = 44.25,\ SD = 30.16$). Considering the standard deviation our results suggest notably large individual differences in all conditions. 
\subsubsection{Errors}
In total 30 navigation errors were recorded. 7 (PM: 2, SP: 5) errors were performed in RL and 23 (PM: 10, SP: 13) in VR. When comparing the navigation methods in both RL and VR the smartphone induced more errors than the paper map. It is important to note that three participants were responsible for 24 (80\%) of all errors which include all of the RL errors and 17 of the VR errors. All remaining 6 errors were performed in VR by different participants.
\subsection{Task Load}
When using the paper map 14 out of 16 participants considered the task load in RL ($M = 30.33,\ SD = 20.77$) to be lower than in VR ($M = 47.25,\ SD = 18.73$). In the smartphone condition all of the participants reported a lower task load in RL ($M = 27.54,\ SD = 16.45$) than in VR ($M = 47.88,\ SD = 18.50$). On average the task load was $\sim36\%$ less in RL than in VR for the paper map and $\sim42\%$ for the smartphone condition.  A Wilcoxon-Signed-Rank test found significance for both the paper map ($p < .001,\ Z = -3.366$) and smartphone condition ($p < .0001,\ Z = -4.170$). When comparing every single subscale of the task load index between RL and VR we always found a lower mean in RL than in VR, regardless of the method. The results suggest that the task load in RL is generally lower than in VR and that the navigation method itself does not seem to have a large impact.

\begin{figure}[t!]
\centering
\includegraphics[width=\columnwidth]{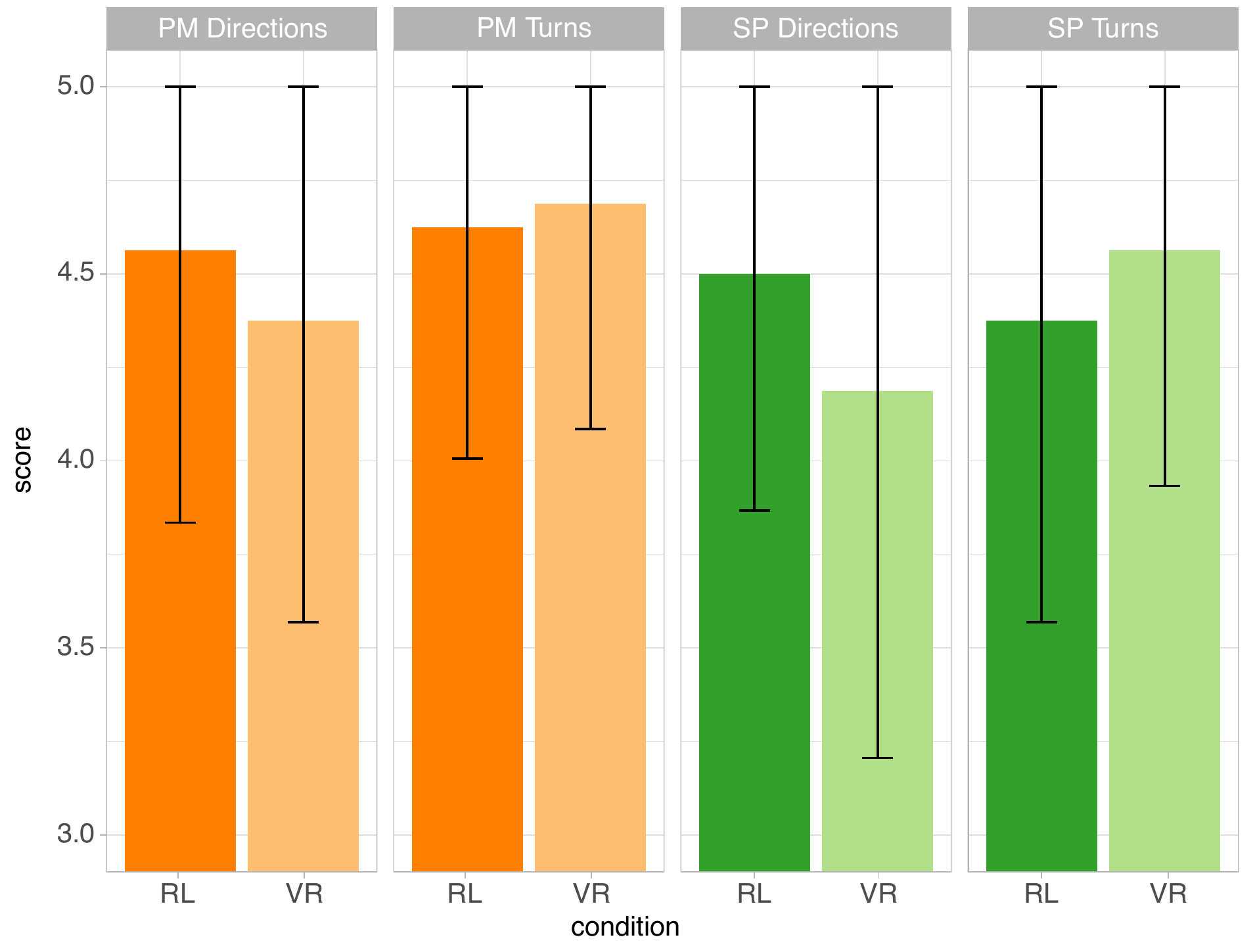}
\caption{Average route recall score for each condition in terms of number of correct turns and number of correct directional changes.}
\label{fig:route_recall}
\end{figure}

\subsection{Spatial Knowledge}
The measurements of the acquired spatial knowledge consisted of a landmark recall and a route recall test. In the landmark recall test participants could achieve a maximum score of 9, while the route recall test was divided into two measurements with a maximum score of 5 points each.

\subsubsection{Landmark Recall Test}
On average the landmark recall was better in RL than in VR for both paper map and smartphone. In the paper map condition the average score decreased from $5.94$ in RL to $5.06$ in VR, while the standard deviation increased slightly from $1.48$ to $1.77$ respectively. We found similar results for the smartphone condition, where the mean score decreased from $5.94$ in RL to $4.75$ in VR, while the standard deviation remained similar (1.18 in both conditions). A pairwise t-test with repeated measures found no significant differences ($t(15) = -1.28, p = 0.2192$) between RL and VR in the paper map condition, but significant differences were revealed when comparing the smartphone conditions ($t(15) = 3.05, p < .01$). When comparing the landmarks themselves we found that in RL all landmarks were recognised at least once, while there were two landmarks in VR that were never recalled by the participants. In VR 14 out of 20 landmarks had a recognition rate below 50\%, while in RL 10 out of 20 had a recognition rate below 50\%.

\subsubsection{Route Recall Test}
No participant scored less than 3 out of 5 possible points for both the number of correct turns ($\#turns$) and the number of correct directional changes ($\#directions$) respectively. The average score in terms of the correct amount of turns slightly increased when comparing RL (PM: 4.63, SP: 4.38) to VR (PM: 4.69, SP: 4.56). In contrast to that the average score for the correct directional changes is fairly higher in RL (PM: 4.56, SP: 4.50) than in VR (PM: 4.38, SP: 4.19). A pairwise Wilcoxon-Signed-Rank test revealed no significant difference for both aspects between RL and VR for paper map (\#turns: $p = 0.766,\ Z = -0.298$, $\#directions$: $p = 0.429,\ Z = -0,791$) and smartphone (\#turns: $p = 0.407,\ Z = -0.829$, \#directions: $p = 0.243,\ Z = -1.167$). Nevertheless it is worth to mention that the smartphone caused a slight decrease in the means with a slightly bigger standard deviation for each aspect in RL and VR compared to the paper map which can be seen in figure \ref{fig:route_recall}.

\section{Discussion \& Conclusion}
\begin{figure*}[t!]
  \includegraphics[width=0.9\textwidth]{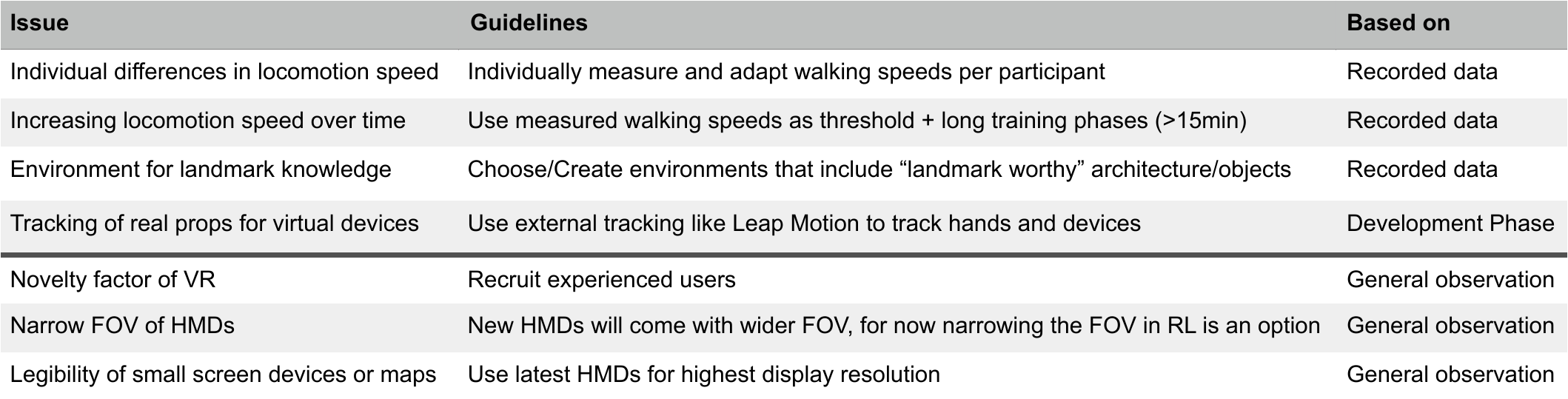}
  \caption{Guidelines for testing navigation methods in VR based on different findings of our study.}
  \label{fig:guidelines}
\end{figure*}

In our study we compare navigation performance, task load and spatial knowledge acquisition using a paper map and smartphone in RL and VR respectively. Overall the results between VR and RL were significantly different in navigation performance, task load and landmark recognition. Route recognition, however, was not significantly different in both RL and VR.  Still  challenges and problems are to overcome until VR  environments can fully replace RL for navigation in an experiment setting. The following discussion will therefore highlight how the locomotion, the VR environment and the navigation method influence participants navigation performance (RQ1). We also present guidelines (see figure \ref{fig:guidelines}) on how we can improve upon these factors and bring VR navigation and RL navigation closer together (RQ2).

\subsection{Locomotion}
WIP proved to be the most natural locomotion technique from the ones that were tested in our pre-study, but that doesn't mean that it doesn't come with drawbacks that need to be discussed. McMahans Framework for Interaction Fidelity Analysis (FIFA) rates WIP slightly above mid-fidelity and found that mid-fidelity interactions mostly perform worse than high- as well as low-fidelity interaction~\cite{McMahan2011,McMahan2016}. Although their implementation differed slightly from ours we would rate our implementation similarly and see the following challenges for pedestrian navigation: The movement times in figure \ref{fig:ct} suggest that participants were overall faster in the VR conditions compared to the RL conditions.  This could be an artefact of the factor translating the up and down movement into forward motion while using our WIP implementation. To better control this in future studies we suggest taking individual walking speed measure in the beginning of both the VR and RL condition. That way experimenters know how much the speed gets in- or decreased in VR on an individual basis.
Another thing we noted was, that participants sped up during the second VR condition as they got used to WIP and tried to exploit the locomotion technique (e.g. jogging, walking through objects). This could be dealt with using the before measured walking speed as a reference and prohibit deviating too much from it.

\subsection{VR Environment}
In comparison between the RL and VR conditions the route recall test did not show significant differences. Due to the short length of the routes it was possible, on both the paper map and the smartphone, to see the whole route at once. So participants could have potentially recalled the image of the route they saw on their device.
Due to the high number of landmarks that we used in our test the landmarks chosen in RL and remodelled in VR, some of them might not have been recognised as such. As most of the streets were regular neighbourhoods we had to deal with a lot of repeating patterns in architecture which made it difficult to generate enough building that qualify as proper landmarks.

As the questionnaire showed, only one participant used VR technology regularly. The rest of the participants have used it either a few times or never before. Being exposed to new technologies often induces a high task load. As our implementation of the WIP technique and the use of our VR props was unique in this setup we expected this to have an influence on the NASA TLX results.

As mentioned in the results we saw a high number of errors were made by only 3 participants. Two of these three experienced strong simulator sickness. However we didn't find any significant correlation in VR between the errors and simulators sickness or errors and familiarity with the environment. 

\subsection{Navigation Methods}
The 110 degrees field of view (FOV) and the low resolution compared to natural vision are the top confounding factors when using a HMD in VR experiments. We found that less FOV results in increased head movement as participants look around or at their navigation device. This potentially decreases the spatial knowledge acquisition and increases the physical demand. The low resolution introduces legibility issues when reading small text on a map or smartphone. Participants had to bring the smartphone screen unnaturally close to their face to read street names when comparing them to the street signs in the VR environment. Especially for smartphone navigation in virtual city environments HMDs with higher display resolution will improve the benefit of street names and other indicators on maps. Most VR implementations use bigger text placed directly in the environment to ensure legibility~\cite{Bowman2003} but due to the realistic form factor of our navigation devices we were bound to a certain maximum text size. 
The results show that in RL participants were a slightly faster with the paper map than with the smartphone when in VR it was the other way round. Additionally in RL participants performed more than double the numbers of errors with the smartphone compared to the paper map whereas in VR the score was a lot more similar. The RL difference could be explained by the simplicity and length of the route. With a longer, more complex route the smartphone might have had the upper hand due to the possibility to zoom into the area of interest but with a short route the map gave a bigger more readable overview in RL. In VR however participants were faster with the smartphone. As figure \ref{fig:navigation_methods} shows participants used real props with the \emph{HTC Vive} controller attached which they could manipulate in VR. This made for a very realistic experience. However it worked better for the smartphone than for the paper map, which could explain the difference in CT and the larger number of errors for the paper map. The controller attached to the smartphone served as a handle to hold the phone and was evenly balanced. The controller attached to the map created an imbalance as now one side of the map was heavier than the other which hindered turning the map to face track-up, which is a common behaviour while navigating~\cite{Montello2006}. Tracking the map and the hands with external hardware could be a solution to resolve this issue. On the smartphone rotation could be easily be performed in VR by using the well known two finger rotate-gesture.  
According to McMahans FIFA~\cite{McMahan2011} our VR paper map and smartphone interaction rates as high-fidelity which is reflected in our general observation of participants using them, especially for the smartphone. The overall touch input on the smartphone seemed to work very well for participants. In prior tests we found that touch accuracy is very high even without highlighting the touch points in VR, as long as the VR model and the RL prop are perfectly aligned. Having the controller attached to the phone in VR also helped a lot. Interestingly the task load was not affected by the navigation method.

VR offers a promising simulation environment to test navigation techniques, but comes with lots of challenges that prevent a true to RL navigation experience. Our research shows that the interaction with two of the most common navigation aids can be successfully implemented using a VE and HMDs in a pedestrian navigation simulation in VR. Still typical problems like field of view and legibility are yet to be solved but will loose importance as the technology develops. Especially the latter has a large influence on the usability of text based navigation devices in VR.
We believe that our results and guidelines will help future researchers to compare multi modal navigation interfaces in VR as well as motivate future work to close the gap between RL field studies and VR lab studies bit by bit. 


\bibliographystyle{ACM-Reference-Format}
\bibliography{main}


\begin{thebibliography}{33}


\ifx \showCODEN    \undefined \def \showCODEN     #1{\unskip}     \fi
\ifx \showDOI      \undefined \def \showDOI       #1{#1}\fi
\ifx \showISBNx    \undefined \def \showISBNx     #1{\unskip}     \fi
\ifx \showISBNxiii \undefined \def \showISBNxiii  #1{\unskip}     \fi
\ifx \showISSN     \undefined \def \showISSN      #1{\unskip}     \fi
\ifx \showLCCN     \undefined \def \showLCCN      #1{\unskip}     \fi
\ifx \shownote     \undefined \def \shownote      #1{#1}          \fi
\ifx \showarticletitle \undefined \def \showarticletitle #1{#1}   \fi
\ifx \showURL      \undefined \def \showURL       {\relax}        \fi
\providecommand\bibfield[2]{#2}
\providecommand\bibinfo[2]{#2}
\providecommand\natexlab[1]{#1}
\providecommand\showeprint[2][]{arXiv:#2}

\bibitem[\protect\citeauthoryear{Aslan, Schwalm, Baus, Kr\"{u}ger, and
  Schwartz}{Aslan et~al\mbox{.}}{2006}]%
        {Aslan:2006}
\bibfield{author}{\bibinfo{person}{Ilhan Aslan}, \bibinfo{person}{Maximilian
  Schwalm}, \bibinfo{person}{J\"{o}rg Baus}, \bibinfo{person}{Antonio
  Kr\"{u}ger}, {and} \bibinfo{person}{Tim Schwartz}.}
  \bibinfo{year}{2006}\natexlab{}.
\newblock \showarticletitle{Acquisition of Spatial Knowledge in Location Aware
  Mobile Pedestrian Navigation Systems}. In
  \bibinfo{booktitle}{\emph{Proceedings of the 8th Conference on Human-computer
  Interaction with Mobile Devices and Services}}
  \emph{(\bibinfo{series}{MobileHCI '06})}. \bibinfo{publisher}{ACM},
  \bibinfo{address}{New York, NY, USA}, \bibinfo{pages}{105--108}.
\newblock
\showISBNx{1-59593-390-5}
\urldef\tempurl%
\url{https://doi.org/10.1145/1152215.1152237}
\showDOI{\tempurl}


\bibitem[\protect\citeauthoryear{Bles, Bos, de~Graaf, Groen, and Wertheim}{Bles
  et~al\mbox{.}}{1998}]%
        {Bles:1998}
\bibfield{author}{\bibinfo{person}{Willem Bles}, \bibinfo{person}{Jelte~E Bos},
  \bibinfo{person}{Bernd de Graaf}, \bibinfo{person}{Eric Groen}, {and}
  \bibinfo{person}{Alexander~H Wertheim}.} \bibinfo{year}{1998}\natexlab{}.
\newblock \showarticletitle{Motion sickness: only one provocative conflict?}
\newblock \bibinfo{journal}{\emph{Brain Research Bulletin}}
  \bibinfo{volume}{47}, \bibinfo{number}{5} (\bibinfo{year}{1998}),
  \bibinfo{pages}{481 -- 487}.
\newblock
\showISSN{0361-9230}
\urldef\tempurl%
\url{https://doi.org/10.1016/S0361-9230(98)00115-4}
\showDOI{\tempurl}


\bibitem[\protect\citeauthoryear{Bliss, Tidwell, and Guest}{Bliss
  et~al\mbox{.}}{1997}]%
        {Bliss1997}
\bibfield{author}{\bibinfo{person}{James~P. Bliss}, \bibinfo{person}{Philip~D.
  Tidwell}, {and} \bibinfo{person}{Michael~A. Guest}.}
  \bibinfo{year}{1997}\natexlab{}.
\newblock \showarticletitle{The Effectiveness of Virtual Reality for
  Administering Spatial Navigation Training to Firefighters}.
\newblock \bibinfo{journal}{\emph{Presence: Teleoper. Virtual Environ.}}
  \bibinfo{volume}{6}, \bibinfo{number}{1} (\bibinfo{date}{Feb.}
  \bibinfo{year}{1997}), \bibinfo{pages}{73--86}.
\newblock
\showISSN{1054-7460}
\urldef\tempurl%
\url{https://doi.org/10.1162/pres.1997.6.1.73}
\showDOI{\tempurl}


\bibitem[\protect\citeauthoryear{Bowman, North, Chen, Polys, Pyla, and
  Yilmaz}{Bowman et~al\mbox{.}}{2003}]%
        {Bowman2003}
\bibfield{author}{\bibinfo{person}{Doug~A. Bowman}, \bibinfo{person}{Chris
  North}, \bibinfo{person}{Jian Chen}, \bibinfo{person}{Nicholas~F. Polys},
  \bibinfo{person}{Pardha~S. Pyla}, {and} \bibinfo{person}{Umur Yilmaz}.}
  \bibinfo{year}{2003}\natexlab{}.
\newblock \showarticletitle{Information-rich Virtual Environments: Theory,
  Tools, and Research Agenda}. In \bibinfo{booktitle}{\emph{Proceedings of the
  ACM Symposium on Virtual Reality Software and Technology}}
  \emph{(\bibinfo{series}{VRST '03})}. \bibinfo{publisher}{ACM},
  \bibinfo{address}{New York, NY, USA}, \bibinfo{pages}{81--90}.
\newblock
\showISBNx{1-58113-569-6}
\urldef\tempurl%
\url{https://doi.org/10.1145/1008653.1008669}
\showDOI{\tempurl}


\bibitem[\protect\citeauthoryear{Burigat and Chittaro}{Burigat and
  Chittaro}{2011}]%
        {Burigat:2011}
\bibfield{author}{\bibinfo{person}{Stefano Burigat} {and} \bibinfo{person}{Luca
  Chittaro}.} \bibinfo{year}{2011}\natexlab{}.
\newblock \showarticletitle{Pedestrian Navigation with Degraded GPS Signal:
  Investigating the Effects of Visualizing Position Uncertainty}. In
  \bibinfo{booktitle}{\emph{Proceedings of the 13th International Conference on
  Human Computer Interaction with Mobile Devices and Services}}
  \emph{(\bibinfo{series}{MobileHCI '11})}. \bibinfo{publisher}{ACM},
  \bibinfo{address}{New York, NY, USA}, \bibinfo{pages}{221--230}.
\newblock
\showISBNx{978-1-4503-0541-9}
\urldef\tempurl%
\url{https://doi.org/10.1145/2037373.2037407}
\showDOI{\tempurl}


\bibitem[\protect\citeauthoryear{Cliburn, Winlock, Rilea, and
  Van~Donsel}{Cliburn et~al\mbox{.}}{2007}]%
        {Cliburn:2007}
\bibfield{author}{\bibinfo{person}{Daniel Cliburn}, \bibinfo{person}{Tess
  Winlock}, \bibinfo{person}{Stacy Rilea}, {and} \bibinfo{person}{Matt
  Van~Donsel}.} \bibinfo{year}{2007}\natexlab{}.
\newblock \showarticletitle{Dynamic Landmark Placement As a Navigation Aid in
  Virtual Worlds}. In \bibinfo{booktitle}{\emph{Proceedings of the 2007 ACM
  Symposium on Virtual Reality Software and Technology}}
  \emph{(\bibinfo{series}{VRST '07})}. \bibinfo{publisher}{ACM},
  \bibinfo{address}{New York, NY, USA}, \bibinfo{pages}{211--214}.
\newblock
\showISBNx{978-1-59593-863-3}
\urldef\tempurl%
\url{https://doi.org/10.1145/1315184.1315225}
\showDOI{\tempurl}


\bibitem[\protect\citeauthoryear{Cushman, Stein, and Duffy}{Cushman
  et~al\mbox{.}}{2008}]%
        {Cushman:2008}
\bibfield{author}{\bibinfo{person}{L.~A. Cushman}, \bibinfo{person}{K. Stein},
  {and} \bibinfo{person}{C.~J. Duffy}.} \bibinfo{year}{2008}\natexlab{}.
\newblock \showarticletitle{{D}etecting navigational deficits in cognitive
  aging and {A}lzheimer disease using virtual reality}.
\newblock \bibinfo{journal}{\emph{Neurology}} \bibinfo{volume}{71},
  \bibinfo{number}{12} (\bibinfo{date}{Sep} \bibinfo{year}{2008}),
  \bibinfo{pages}{888--895}.
\newblock


\bibitem[\protect\citeauthoryear{Dancu, Fourgeaud, Obaid, Fjeld, and
  Elmqvist}{Dancu et~al\mbox{.}}{2015}]%
        {Dancu:2015}
\bibfield{author}{\bibinfo{person}{Alexandru Dancu},
  \bibinfo{person}{Micka\"{e}l Fourgeaud}, \bibinfo{person}{Mohammad Obaid},
  \bibinfo{person}{Morten Fjeld}, {and} \bibinfo{person}{Niklas Elmqvist}.}
  \bibinfo{year}{2015}\natexlab{}.
\newblock \showarticletitle{Map Navigation Using a Wearable Mid-air Display}.
  In \bibinfo{booktitle}{\emph{Proceedings of the 17th International Conference
  on Human-Computer Interaction with Mobile Devices and Services}}
  \emph{(\bibinfo{series}{MobileHCI '15})}. \bibinfo{publisher}{ACM},
  \bibinfo{address}{New York, NY, USA}, \bibinfo{pages}{71--76}.
\newblock
\showISBNx{978-1-4503-3652-9}
\urldef\tempurl%
\url{https://doi.org/10.1145/2785830.2785876}
\showDOI{\tempurl}


\bibitem[\protect\citeauthoryear{Darken, Cockayne, and Carmein}{Darken
  et~al\mbox{.}}{1997}]%
        {Darken:1997}
\bibfield{author}{\bibinfo{person}{Rudolph~P. Darken},
  \bibinfo{person}{William~R. Cockayne}, {and} \bibinfo{person}{David
  Carmein}.} \bibinfo{year}{1997}\natexlab{}.
\newblock \showarticletitle{The Omni-directional Treadmill: A Locomotion Device
  for Virtual Worlds}. In \bibinfo{booktitle}{\emph{Proceedings of the 10th
  Annual ACM Symposium on User Interface Software and Technology}}
  \emph{(\bibinfo{series}{UIST '97})}. \bibinfo{publisher}{ACM},
  \bibinfo{address}{New York, NY, USA}, \bibinfo{pages}{213--221}.
\newblock
\showISBNx{0-89791-881-9}
\urldef\tempurl%
\url{https://doi.org/10.1145/263407.263550}
\showDOI{\tempurl}


\bibitem[\protect\citeauthoryear{Deb, Carruth, Sween, Strawderman, and
  Garrison}{Deb et~al\mbox{.}}{2017}]%
        {Deb:2017}
\bibfield{author}{\bibinfo{person}{Shuchisnigdha Deb},
  \bibinfo{person}{Daniel~W. Carruth}, \bibinfo{person}{Richard Sween},
  \bibinfo{person}{Lesley Strawderman}, {and} \bibinfo{person}{Teena~M.
  Garrison}.} \bibinfo{year}{2017}\natexlab{}.
\newblock \showarticletitle{Efficacy of virtual reality in pedestrian safety
  research}.
\newblock \bibinfo{journal}{\emph{Applied Ergonomics}}  \bibinfo{volume}{65}
  (\bibinfo{year}{2017}), \bibinfo{pages}{449 -- 460}.
\newblock
\showISSN{0003-6870}
\urldef\tempurl%
\url{https://doi.org/10.1016/j.apergo.2017.03.007}
\showDOI{\tempurl}


\bibitem[\protect\citeauthoryear{Delikostidis, Fritze, Fechner, and
  Kray}{Delikostidis et~al\mbox{.}}{2015}]%
        {Delikostidis:2015}
\bibfield{author}{\bibinfo{person}{Ioannis Delikostidis},
  \bibinfo{person}{Holger Fritze}, \bibinfo{person}{Thore Fechner}, {and}
  \bibinfo{person}{Christian Kray}.} \bibinfo{year}{2015}\natexlab{}.
\newblock \bibinfo{booktitle}{\emph{Bridging the Gap Between Field- and
  Lab-Based User Studies for Location-Based Services}}.
\newblock \bibinfo{publisher}{Springer International Publishing},
  \bibinfo{address}{Cham}, \bibinfo{pages}{257--271}.
\newblock
\showISBNx{978-3-319-11879-6}
\urldef\tempurl%
\url{https://doi.org/10.1007/978-3-319-11879-6_18}
\showDOI{\tempurl}


\bibitem[\protect\citeauthoryear{Farrel, Arnold, Pettifer, Adams, Graham, and
  MacManamon}{Farrel et~al\mbox{.}}{2003}]%
        {Farrel:2003}
\bibfield{author}{\bibinfo{person}{Martin~J. Farrel}, \bibinfo{person}{Paul
  Arnold}, \bibinfo{person}{Steve Pettifer}, \bibinfo{person}{Jessica Adams},
  \bibinfo{person}{Tom Graham}, {and} \bibinfo{person}{Michael MacManamon}.}
  \bibinfo{year}{2003}\natexlab{}.
\newblock \showarticletitle{Transfer of Route Learning From Virtual to Real
  Environments.}
\newblock \bibinfo{journal}{\emph{Journal of Experimental Psychology: Applied}}
  \bibinfo{volume}{9}, \bibinfo{number}{4} (\bibinfo{year}{2003}),
  \bibinfo{pages}{219--227}.
\newblock
\urldef\tempurl%
\url{https://doi.org/10.1037/1076-898X.9.4.219}
\showDOI{\tempurl}
\showeprint{https://doi.org/10.1037/1076-898X.9.4.219}
\newblock
\shownote{PMID: 14664673.}


\bibitem[\protect\citeauthoryear{Feldstein, Dietrich, Milinkovic, and
  Bengler}{Feldstein et~al\mbox{.}}{2016}]%
        {Feldstein:2016}
\bibfield{author}{\bibinfo{person}{Ilja Feldstein}, \bibinfo{person}{Andr\'{e}
  Dietrich}, \bibinfo{person}{Sasha Milinkovic}, {and} \bibinfo{person}{Klaus
  Bengler}.} \bibinfo{year}{2016}\natexlab{}.
\newblock \showarticletitle{A Pedestrian Simulator for Urban Crossing
  Scenarios}.
\newblock \bibinfo{journal}{\emph{IFAC-PapersOnLine}} \bibinfo{volume}{49},
  \bibinfo{number}{19} (\bibinfo{year}{2016}), \bibinfo{pages}{239 -- 244}.
\newblock
\showISSN{2405-8963}
\urldef\tempurl%
\url{https://doi.org/10.1016/j.ifacol.2016.10.531}
\showDOI{\tempurl}
\newblock
\shownote{13th IFAC Symposium on Analysis, Design, and Evaluation
  ofHuman-Machine Systems HMS 2016.}


\bibitem[\protect\citeauthoryear{Frommer}{Frommer}{2017}]%
        {Frommer:2017}
\bibfield{author}{\bibinfo{person}{Dan Frommer}.}
  \bibinfo{year}{2017}\natexlab{}.
\newblock \bibinfo{title}{These are the 10 most popular mobile apps in
  America}.
\newblock \bibinfo{howpublished}{Blog}.
\newblock
\newblock
\shownote{Retrieved August 28, 2017 from
  \url{http://www.recode.net/2017/8/24/16197218/top-10-mobile-apps-2017-comscore-chart-facebook-google}.}


\bibitem[\protect\citeauthoryear{Gordon, Lucy, and Michael}{Gordon
  et~al\mbox{.}}{2003}]%
        {Simpson:2003}
\bibfield{author}{\bibinfo{person}{Simpson Gordon}, \bibinfo{person}{Johnston
  Lucy}, {and} \bibinfo{person}{Richardson Michael}.}
  \bibinfo{year}{2003}\natexlab{}.
\newblock \showarticletitle{An investigation of road crossing in a virtual
  environment}.
\newblock \bibinfo{journal}{\emph{Accident Analysis \& Prevention}}
  \bibinfo{volume}{35}, \bibinfo{number}{5} (\bibinfo{year}{2003}),
  \bibinfo{pages}{787 -- 796}.
\newblock
\showISSN{0001-4575}
\urldef\tempurl%
\url{https://doi.org/10.1016/S0001-4575(02)00081-7}
\showDOI{\tempurl}


\bibitem[\protect\citeauthoryear{Hart and Staveland}{Hart and
  Staveland}{1988}]%
        {Hart1988}
\bibfield{author}{\bibinfo{person}{Sandra~G. Hart} {and}
  \bibinfo{person}{Lowell~E. Staveland}.} \bibinfo{year}{1988}\natexlab{}.
\newblock \showarticletitle{Development of NASA-TLX (Task Load Index): Results
  of Empirical and Theoretical Research}.
\newblock In \bibinfo{booktitle}{\emph{Human Mental Workload}},
  \bibfield{editor}{\bibinfo{person}{Peter~A. Hancock} {and}
  \bibinfo{person}{Najmedin Meshkati}} (Eds.). \bibinfo{series}{Advances in
  Psychology}, Vol.~\bibinfo{volume}{52}. \bibinfo{publisher}{North-Holland},
  \bibinfo{pages}{139 -- 183}.
\newblock
\showISSN{0166-4115}
\urldef\tempurl%
\url{https://doi.org/10.1016/S0166-4115(08)62386-9}
\showDOI{\tempurl}


\bibitem[\protect\citeauthoryear{Kjeldskov, Skov, Als, and H{\o}egh}{Kjeldskov
  et~al\mbox{.}}{2004}]%
        {Kjeldskov:2004}
\bibfield{author}{\bibinfo{person}{Jesper Kjeldskov},
  \bibinfo{person}{Mikael~B. Skov}, \bibinfo{person}{Benedikte~S. Als}, {and}
  \bibinfo{person}{Rune~T. H{\o}egh}.} \bibinfo{year}{2004}\natexlab{}.
\newblock \showarticletitle{Is It Worth the Hassle? Exploring the Added Value
  of Evaluating the Usability of Context-Aware Mobile Systems in the Field}. In
  \bibinfo{booktitle}{\emph{Mobile Human-Computer Interaction - MobileHCI
  2004}}, \bibfield{editor}{\bibinfo{person}{Stephen Brewster} {and}
  \bibinfo{person}{Mark Dunlop}} (Eds.). \bibinfo{publisher}{Springer Berlin
  Heidelberg}, \bibinfo{address}{Berlin, Heidelberg}.
\newblock
\showISBNx{978-3-540-28637-0}


\bibitem[\protect\citeauthoryear{Kluss, Marsh, Zetzsche, and Schill}{Kluss
  et~al\mbox{.}}{2015}]%
        {Kluss:2015}
\bibfield{author}{\bibinfo{person}{Thorsten Kluss}, \bibinfo{person}{William~E.
  Marsh}, \bibinfo{person}{Christoph Zetzsche}, {and} \bibinfo{person}{Kerstin
  Schill}.} \bibinfo{year}{2015}\natexlab{}.
\newblock \showarticletitle{Representation of impossible worlds in the
  cognitive map}.
\newblock \bibinfo{journal}{\emph{Cognitive Processing}} \bibinfo{volume}{16},
  \bibinfo{number}{1} (\bibinfo{date}{01 Sep} \bibinfo{year}{2015}),
  \bibinfo{pages}{271--276}.
\newblock
\showISSN{1612-4790}
\urldef\tempurl%
\url{https://doi.org/10.1007/s10339-015-0705-x}
\showDOI{\tempurl}


\bibitem[\protect\citeauthoryear{Lapeyre, Hourlier, Servantie, N'Kaoua, and
  Sauz\'eon}{Lapeyre et~al\mbox{.}}{2011}]%
        {Lapeyre:2011}
\bibfield{author}{\bibinfo{person}{Brigitte Lapeyre}, \bibinfo{person}{Sylvain
  Hourlier}, \bibinfo{person}{Xavier Servantie}, \bibinfo{person}{Bernard
  N'Kaoua}, {and} \bibinfo{person}{H\'el\`ene Sauz\'eon}.}
  \bibinfo{year}{2011}\natexlab{}.
\newblock \showarticletitle{Using the Landmark-Route-Survey Framework to
  Evaluate Spatial Knowledge Obtained From Synthetic Vision Systems}.
\newblock \bibinfo{journal}{\emph{Human Factors}} \bibinfo{volume}{53},
  \bibinfo{number}{6} (\bibinfo{year}{2011}), \bibinfo{pages}{647--661}.
\newblock
\urldef\tempurl%
\url{https://doi.org/10.1177/0018720811421171}
\showDOI{\tempurl}


\bibitem[\protect\citeauthoryear{Lehsing, Feldstein, Dietrich, and
  Bengler}{Lehsing et~al\mbox{.}}{2016}]%
        {Lehsing:2016}
\bibfield{author}{\bibinfo{person}{Christian Lehsing}, \bibinfo{person}{Ilja
  Feldstein}, \bibinfo{person}{Andr\'e Dietrich}, {and} \bibinfo{person}{Klaus
  Bengler}.} \bibinfo{year}{2016}\natexlab{}.
\newblock \showarticletitle{Pedestrian Simulator for Traffic Research - State
  of the Art and Future of a Motion Lab}.
\newblock  (\bibinfo{date}{06} \bibinfo{year}{2016}).
\newblock


\bibitem[\protect\citeauthoryear{Maffei, Masullo, Sorrentino, and
  Gabriele}{Maffei et~al\mbox{.}}{2014}]%
        {Maffei:2014}
\bibfield{author}{\bibinfo{person}{Luigi Maffei}, \bibinfo{person}{Massimiliano
  Masullo}, \bibinfo{person}{Francesco Sorrentino}, {and}
  \bibinfo{person}{Maria Gabriele}.} \bibinfo{year}{2014}\natexlab{}.
\newblock \showarticletitle{Preliminary studies on the relation between the
  audio-visual cues' perception and the approaching speed of electric
  vehicles}.
\newblock   \bibinfo{volume}{20} (\bibinfo{date}{04} \bibinfo{year}{2014}),
  \bibinfo{pages}{1--9}.
\newblock


\bibitem[\protect\citeauthoryear{McMahan}{McMahan}{2011}]%
        {McMahan2011}
\bibfield{author}{\bibinfo{person}{Ryan~P. McMahan}.}
  \bibinfo{year}{2011}\natexlab{}.
\newblock \emph{\bibinfo{title}{Exploring the Effects of Higher-Fidelity
  Display and Interaction for Virtual Reality Games}}.
\newblock {PhD} dissertation. \bibinfo{school}{Virginia Tech}.
\newblock


\bibitem[\protect\citeauthoryear{McMahan, Lai, and Pal}{McMahan
  et~al\mbox{.}}{2016}]%
        {McMahan2016}
\bibfield{author}{\bibinfo{person}{Ryan~P. McMahan}, \bibinfo{person}{Chengyuan
  Lai}, {and} \bibinfo{person}{Swaroop~K. Pal}.}
  \bibinfo{year}{2016}\natexlab{}.
\newblock \showarticletitle{Interaction Fidelity: The Uncanny Valley of Virtual
  Reality Interactions}. In \bibinfo{booktitle}{\emph{Virtual, Augmented and
  Mixed Reality - 8th International Conference, {VAMR} 2016, Held as Part of
  {HCI} International 2016, Toronto, Canada, July 17-22, 2016. Proceedings}}.
  \bibinfo{pages}{59--70}.
\newblock
\urldef\tempurl%
\url{https://doi.org/10.1007/978-3-319-39907-2\_6}
\showDOI{\tempurl}


\bibitem[\protect\citeauthoryear{Pielot and Boll}{Pielot and Boll}{2010}]%
        {Pielot:2010}
\bibfield{author}{\bibinfo{person}{Martin Pielot} {and}
  \bibinfo{person}{Susanne Boll}.} \bibinfo{year}{2010}\natexlab{}.
\newblock \showarticletitle{Tactile Wayfinder: Comparison of Tactile Waypoint
  Navigation with Commercial Pedestrian Navigation Systems}. In
  \bibinfo{booktitle}{\emph{Pervasive Computing}},
  \bibfield{editor}{\bibinfo{person}{Patrik Flor{\'e}en},
  \bibinfo{person}{Antonio Kr{\"u}ger}, {and} \bibinfo{person}{Mirjana
  Spasojevic}} (Eds.). \bibinfo{publisher}{Springer Berlin Heidelberg},
  \bibinfo{address}{Berlin, Heidelberg}, \bibinfo{pages}{76--93}.
\newblock
\showISBNx{978-3-642-12654-3}


\bibitem[\protect\citeauthoryear{R~Montello and Sas}{R~Montello and
  Sas}{2006}]%
        {Montello2006}
\bibfield{author}{\bibinfo{person}{Daniel R~Montello} {and}
  \bibinfo{person}{Corina Sas}.} \bibinfo{year}{2006}\natexlab{}.
\newblock \showarticletitle{Human Factors of Wayfinding in Navigation}.
\newblock \bibinfo{journal}{\emph{International Encyclopedia of Ergonomics and
  Human Factors}} (\bibinfo{date}{03} \bibinfo{year}{2006}).
\newblock
\urldef\tempurl%
\url{https://doi.org/10.1201/9780849375477.ch394}
\showDOI{\tempurl}


\bibitem[\protect\citeauthoryear{Richardson, Montello, and Hegarty}{Richardson
  et~al\mbox{.}}{1999}]%
        {Richardson:1999}
\bibfield{author}{\bibinfo{person}{Anthony~E. Richardson},
  \bibinfo{person}{Daniel~R. Montello}, {and} \bibinfo{person}{Mary Hegarty}.}
  \bibinfo{year}{1999}\natexlab{}.
\newblock \showarticletitle{Spatial knowledge acquisition from maps and from
  navigation in real and virtual environments}.
\newblock \bibinfo{journal}{\emph{Memory {\&} Cognition}} \bibinfo{volume}{27},
  \bibinfo{number}{4} (\bibinfo{date}{01 Jul} \bibinfo{year}{1999}),
  \bibinfo{pages}{741--750}.
\newblock
\showISSN{1532-5946}
\urldef\tempurl%
\url{https://doi.org/10.3758/BF03211566}
\showDOI{\tempurl}


\bibitem[\protect\citeauthoryear{Schirmer, Hartmann, Bertel, and
  Echtler}{Schirmer et~al\mbox{.}}{2015}]%
        {Schirmer:2015}
\bibfield{author}{\bibinfo{person}{Maximilian Schirmer},
  \bibinfo{person}{Johannes Hartmann}, \bibinfo{person}{Sven Bertel}, {and}
  \bibinfo{person}{Florian Echtler}.} \bibinfo{year}{2015}\natexlab{}.
\newblock \showarticletitle{Shoe Me the Way: A Shoe-Based Tactile Interface for
  Eyes-Free Urban Navigation}. In \bibinfo{booktitle}{\emph{Proceedings of the
  17th International Conference on Human-Computer Interaction with Mobile
  Devices and Services}} \emph{(\bibinfo{series}{MobileHCI '15})}.
  \bibinfo{publisher}{ACM}, \bibinfo{address}{New York, NY, USA},
  \bibinfo{pages}{327--336}.
\newblock
\showISBNx{978-1-4503-3652-9}
\urldef\tempurl%
\url{https://doi.org/10.1145/2785830.2785832}
\showDOI{\tempurl}


\bibitem[\protect\citeauthoryear{Schneegass, Alt, Scheible, Schmidt, and
  Su}{Schneegass et~al\mbox{.}}{2014}]%
        {Schneegass:2014}
\bibfield{author}{\bibinfo{person}{Stefan Schneegass}, \bibinfo{person}{Florian
  Alt}, \bibinfo{person}{J\"{u}rgen Scheible}, \bibinfo{person}{Albrecht
  Schmidt}, {and} \bibinfo{person}{Haifeng Su}.}
  \bibinfo{year}{2014}\natexlab{}.
\newblock \showarticletitle{Midair Displays: Exploring the Concept of
  Free-floating Public Displays}. In \bibinfo{booktitle}{\emph{CHI '14 Extended
  Abstracts on Human Factors in Computing Systems}} \emph{(\bibinfo{series}{CHI
  EA '14})}. \bibinfo{publisher}{ACM}, \bibinfo{address}{New York, NY, USA},
  \bibinfo{pages}{2035--2040}.
\newblock
\showISBNx{978-1-4503-2474-8}
\urldef\tempurl%
\url{https://doi.org/10.1145/2559206.2581190}
\showDOI{\tempurl}


\bibitem[\protect\citeauthoryear{Suma, Azmandian, Grechkin, Phan, and
  Bolas}{Suma et~al\mbox{.}}{2015}]%
        {Suma:2015}
\bibfield{author}{\bibinfo{person}{Evan~A. Suma}, \bibinfo{person}{Mahdi
  Azmandian}, \bibinfo{person}{Timofey Grechkin}, \bibinfo{person}{Thai Phan},
  {and} \bibinfo{person}{Mark Bolas}.} \bibinfo{year}{2015}\natexlab{}.
\newblock \showarticletitle{Making Small Spaces Feel Large: Infinite Walking in
  Virtual Reality}. In \bibinfo{booktitle}{\emph{ACM SIGGRAPH 2015 Emerging
  Technologies}} \emph{(\bibinfo{series}{SIGGRAPH '15})}.
  \bibinfo{publisher}{ACM}, \bibinfo{address}{New York, NY, USA}, Article
  \bibinfo{articleno}{16}, \bibinfo{numpages}{1}~pages.
\newblock
\showISBNx{978-1-4503-3635-2}
\urldef\tempurl%
\url{https://doi.org/10.1145/2782782.2792496}
\showDOI{\tempurl}


\bibitem[\protect\citeauthoryear{Szymczak, Rassmus-Gr\"{o}hn, Magnusson, and
  Hedvall}{Szymczak et~al\mbox{.}}{2012}]%
        {Szymczak:2012}
\bibfield{author}{\bibinfo{person}{Delphine Szymczak}, \bibinfo{person}{Kirsten
  Rassmus-Gr\"{o}hn}, \bibinfo{person}{Charlotte Magnusson}, {and}
  \bibinfo{person}{Per-Olof Hedvall}.} \bibinfo{year}{2012}\natexlab{}.
\newblock \showarticletitle{A Real-world Study of an Audio-tactile Tourist
  Guide}. In \bibinfo{booktitle}{\emph{Proceedings of the 14th International
  Conference on Human-computer Interaction with Mobile Devices and Services}}
  \emph{(\bibinfo{series}{MobileHCI '12})}. \bibinfo{publisher}{ACM},
  \bibinfo{address}{New York, NY, USA}, \bibinfo{pages}{335--344}.
\newblock
\showISBNx{978-1-4503-1105-2}
\urldef\tempurl%
\url{https://doi.org/10.1145/2371574.2371627}
\showDOI{\tempurl}


\bibitem[\protect\citeauthoryear{Wallet, Sauz{\'e}on, Rodrigues, and
  N'Kaoua}{Wallet et~al\mbox{.}}{2008}]%
        {Wallet:2008}
\bibfield{author}{\bibinfo{person}{Gr{\'e}gory Wallet},
  \bibinfo{person}{H{\'e}l\`{e}ne Sauz{\'e}on}, \bibinfo{person}{J{\'e}r\^{o}me
  Rodrigues}, {and} \bibinfo{person}{Bernard N'Kaoua}.}
  \bibinfo{year}{2008}\natexlab{}.
\newblock \showarticletitle{Use of Virtual Reality for Spatial Knowledge
  Transfer: Effects of Passive/Active Exploration Mode in Simple and Complex
  Routes for Three Different Recall Tasks}. In
  \bibinfo{booktitle}{\emph{Proceedings of the 2008 ACM Symposium on Virtual
  Reality Software and Technology}} \emph{(\bibinfo{series}{VRST '08})}.
  \bibinfo{publisher}{ACM}, \bibinfo{address}{New York, NY, USA},
  \bibinfo{pages}{175--178}.
\newblock
\showISBNx{978-1-59593-951-7}
\urldef\tempurl%
\url{https://doi.org/10.1145/1450579.1450616}
\showDOI{\tempurl}


\bibitem[\protect\citeauthoryear{Walther-Franks, Wenig, Smeddinck, and
  Malaka}{Walther-Franks et~al\mbox{.}}{2013}]%
        {Walther-Franks:2013}
\bibfield{author}{\bibinfo{person}{Benjamin Walther-Franks},
  \bibinfo{person}{Dirk Wenig}, \bibinfo{person}{Jan Smeddinck}, {and}
  \bibinfo{person}{Rainer Malaka}.} \bibinfo{year}{2013}\natexlab{}.
\newblock \showarticletitle{Suspended Walking: A Physical Locomotion Interface
  for Virtual Reality}. In \bibinfo{booktitle}{\emph{Entertainment Computing --
  ICEC 2013}}, \bibfield{editor}{\bibinfo{person}{Junia~C. Anacleto},
  \bibinfo{person}{Esteban W.~G. Clua}, \bibinfo{person}{Flavio S.~Correa
  da~Silva}, \bibinfo{person}{Sidney Fels}, {and} \bibinfo{person}{Hyun~S.
  Yang}} (Eds.). \bibinfo{publisher}{Springer Berlin Heidelberg},
  \bibinfo{address}{Berlin, Heidelberg}, \bibinfo{pages}{185--188}.
\newblock
\showISBNx{978-3-642-41106-9}


\bibitem[\protect\citeauthoryear{Wenig, Sch\"{o}ning, Hecht, and Malaka}{Wenig
  et~al\mbox{.}}{2015}]%
        {Wenig:2015}
\bibfield{author}{\bibinfo{person}{Dirk Wenig}, \bibinfo{person}{Johannes
  Sch\"{o}ning}, \bibinfo{person}{Brent Hecht}, {and} \bibinfo{person}{Rainer
  Malaka}.} \bibinfo{year}{2015}\natexlab{}.
\newblock \showarticletitle{StripeMaps: Improving Map-based Pedestrian
  Navigation for Smartwatches}. In \bibinfo{booktitle}{\emph{Proceedings of the
  17th International Conference on Human-Computer Interaction with Mobile
  Devices and Services}} \emph{(\bibinfo{series}{MobileHCI '15})}.
  \bibinfo{publisher}{ACM}, \bibinfo{address}{New York, NY, USA},
  \bibinfo{pages}{52--62}.
\newblock
\showISBNx{978-1-4503-3652-9}
\urldef\tempurl%
\url{https://doi.org/10.1145/2785830.2785862}
\showDOI{\tempurl}


\end{thebibliography}


\end{document}